# Implementation of Function Point Analysis in Measuring the Volume Estimation of Software System in Object Oriented and Structural Model of Academic System

Dian Pratiwi
Trisakti University
Jl. Kyai Tapa No.1
Jakarta, 15000, Indonesia

## ABSTRACT
In the software development required a fidelity and accuracy in determining the size or value of the software to fit the operation is executed. Various methods of calculation has been widely applied to estimate the size, and one of them is by using the method of Function Point Analysis (FPA). The method is then applied by author to measure the complexity of an academic information system by using the two modeling approaches, namely object oriented and structural models. Measurements in this paper consists of several stages, namely describing the information system that will be built into the UML models and structured. Then the model is analyzed by calculating Crude Function Points (CRP), Relative Complexity Adjustment Factor (RCAF), and then calculate its function point. From the result of a calculation using the FPA to the academic system software development, FP values of object oriented model obtained for 174,64 and the FP value of structured models for 180,93. The result of function point that will be used by developers in determining the price and cost of software systems to be built.

## General Terms
Volume Software Estimation

## 1. INTRODUCTION
The success of a software project is determined by various factors that are related each other in the project. A project will be called a success if all the requirements can be fulfilled, the cost is not excessive (overflow), did not pass through the schedule (deadline) that has been planned. This can be done either with a way of estimating the size of software volume systems that will provide the precision of the complexity and value of the product prices on the software project. So that the developer can plan resources, cost, and duration required precisely to build a piece of software.

In this paper, we will discuss the use of Function Point Analysis (FPA) method to measuring the volume estimated of a software system, namely the academic system an educational institution which will be compared on the estimation of the model using the Unified Modeling Language (UML) based object model with the Data Flow Diagram (DFD) based structured model.

## 2. FUNCTION POINT ANALYSIS (FPA)
FPA method is a part of The FSM (Functional Size Measurement) method was first introduced by Albrecht in 1979 as a method for measuring the amount of complexity and functionality in a software project. In the FPA procedure there are a variety of transactions, comprising the incoming and outgoing data to be processed on the system. Each transaction in the FPA will be mapped to the following models

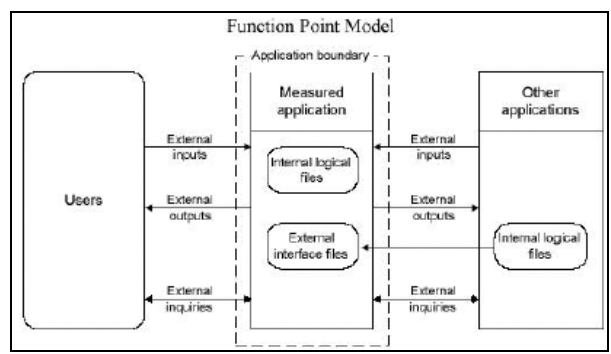

**Fig 1: Function Point Method [2]**

In the picture above can be seen, The Function Point Model consists of [3] :

1. External Input (EI)
   Functions that move data into the application without presenting data manipulation.
2. External Output (EO)
   Functions that move data to user and presents some data manipulation.
3. External Inquiries (EQ)
   Functions that move data to user without presenting data manipulation.
4. Internal Logical Files (ILF)
   The logic in the form of fixed data managed by the application through the use of External Input (EI)
5. External Interface Files (EIF)
   The logic in the form of fixed data used by the application but did not run in it

In the computation phase, each transaction is sorted by the amount of data that they use. Logical transaction or file sorted based on the entities (called RET or Referenced Entity Types) and attributes (called DET or Data Entity Types). Functional transaction sorted based on the attribute numbers (DET), which moved out of the line and the numbering of logical transaction references. Then the whole categorized into 'low', 'average', or 'high' which each category given value as the value of Function Point (FP) [1].

The stages are there in determining the function point is [5]:

1. Calculating The CFP (Crude Function Points)
   The number of functional components of the system were first identified and followed to evaluated the complexity of quantization weight of each component. Weighting was the summed and become the number of CFP. CFP calculation involves five types of software system components following :
   - The number of input applications
   - The number of output applications
   - The number of online query applications.
   - This application related to query against the data stored
   - The number of logical files or tables which involved
   - The number of external interfaces
     An output or input interface that can be connected to the computer through data communications, CDs, floppy disks, etc.

   Then given a weighting factor to each of the above components based on its complexity. The table below is an example of the weighting blank :

**Table 1. Blank of CFP Calculation [6]**

| Software System Components | Level of Complexity | | | | | | | | | Sum of CFP |
|---|---|---|---|---|---|---|---|---|---|---|
| | Low | | | Average | | | High | | | |
| | Count | Weighting Factor | Point | Count | Weighting Factor | Point | Count | Weighting Factor | Point | |
| | A | B | C = AxB | D | E | F = DxE | G | H | I = GxH | J = C+F+I |
| Input | | 3 | | | 4 | | | 6 | | |
| Output | | 4 | | | 5 | | | 7 | | |
| Online Query | | 3 | | | 4 | | | 6 | | |
| File Logic | | 7 | | | 10 | | | 15 | | |
| External Interface | | 5 | | | 7 | | | 10 | | |
| Sum of CFP | | | | | | | | | | |

2. Calculating the complexity of transcription factors of RCAF (Relative Complexity Adjustment Factor) for the project.
   RCAF is to calculate the complexity assessment of software system from several characteristics of subject. Rating scale from 0 to 5 is given to each subject that most affect the development effort required. Example of RCAF assessment form can be seen as follows :

**Table 2. RCAF Assessment Form [6]**

| No. | Subject | Value | | | | | |
|---|---|---|---|---|---|---|---|
| 1 | The level of recovery reliability complexity | 0 | 1 | 2 | 3 | 4 | 5 |
| 2 | The level of data communication complexity | 0 | 1 | 2 | 3 | 4 | 5 |
| 3 | The level of distributed processing complexity | 0 | 1 | 2 | 3 | 4 | 5 |
| 4 | Level of the need for performance complexity | 0 | 1 | 2 | 3 | 4 | 5 |
| 5 | The level of operating environment demand | 0 | 1 | 2 | 3 | 4 | 5 |
| 6 | The level of developer knowledge needs | 0 | 1 | 2 | 3 | 4 | 5 |
| 7 | The level of updating the master file complexity | 0 | 1 | 2 | 3 | 4 | 5 |
| 8 | The level of installation complexity | 0 | 1 | 2 | 3 | 4 | 5 |
| 9 | The level of input, output, online query and file application complexity | 0 | 1 | 2 | 3 | 4 | 5 |
| 10 | The level of data processing complexity | 0 | 1 | 2 | 3 | 4 | 5 |
| 11 | The improbability level of reuse code | 0 | 1 | 2 | 3 | 4 | 5 |
| 12 | The level of customer organization variation | 0 | 1 | 2 | 3 | 4 | 5 |
| 13 | The extent of possible changes | 0 | 1 | 2 | 3 | 4 | 5 |
| 14 | Level of the ease of use demand | 0 | 1 | 2 | 3 | 4 | 5 |
| | Total = RCAF | | | | | | |

3. Calculating Function Points by the formula [6]
$$FP = CFP \times (0.65 + 0.01 \times RCAF) \qquad (1)$$

## 3. OBJECT ORIENTED MODEL

Object oriented model that is intended in this paper is a model using UML (Unified Modeling Language) which has several UML diagrams as a form of visualization, specification, and documentation of software systems. The division is in the UML diagrams such as use-case diagram, class diagram, sequence diagram, activity diagram, state-chart diagram, collaboration diagram, component diagram, and deployment diagram. Of the eight diagrams in the UML diagram, which will be used in this paper is only two, in which the behavioral models are use-case diagram, and structural models is a class-diagram.

Use-case diagram is a diagram that shows the connection between the actor, the subject (or system), and use-case. This diagram describes the system in a box and use-case in an oval shape. A use-case represents the functionality of the system, and the actor who access it. While the class diagram is a diagram that shows a collection of the declaration of the model elements, such as classes, class content and relationships between classes [4].

## 4. STRUCTURAL MODEL

A structured approach in this paper is to use a model of Data Flow Diagram (DFD) and Entity Relationship Diagram (ERD). DFD is a diagram to describe the system as a network of functional process that connected to each other by the data flow. While the ERD is a model from the logical structure of database.

In DFD, there are some important components (according Yourdan and De Marco) which is component as terminators in a box of a source or destination, the components in the form of a circle of as a description of the activities that will be/are being implemented consisting of input and output, data store component in the form of two horizontal lines are lined up as an image of computerized storage, and data flow components depicted in the form of an arrow as data transfer from one section to another system.

In ERD, the modeling made up of several components, namely the entities, attributes, relationships, cardinality ratio and constraint participant. An entity is an object that will store, produce information. Attributes are the characteristics of the entity or relationship that provides a detailed explanation of the entity or relationship. Relationship is the relationship between one or more entities. The cardinality ratio is a limit to the number of connection one entity to another entity. While the constraint participant clarify whether the existence of an entity depends on its relationship with other entities

## 5. PROCEDURE & IMPLEMENTATION

In this study, will be built a system of academic services for students at a university, where the system will provide services such as filling the Study Plan Card or KRS online, check the schedule of lectures, and check the courses value.

### 5.1 UML Modeling
- Use-Case Diagram

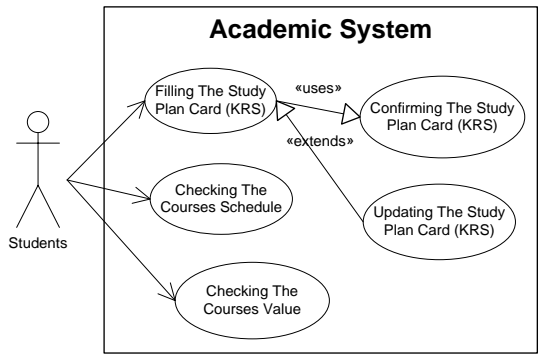

**Fig 2: Use Case Diagram of Academic System**

- Class Diagram

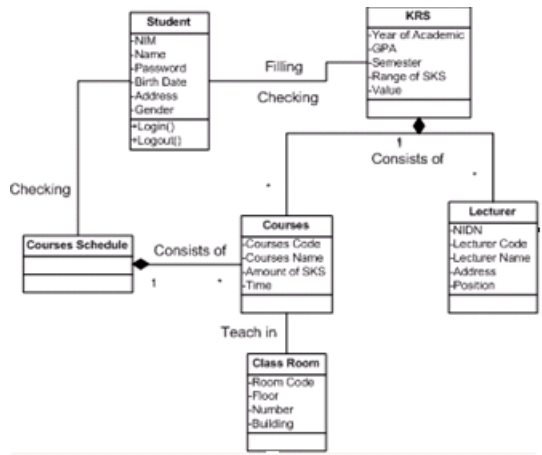

**Fig 3: Class Diagram of Academic System**

- Function Point Calculation

Function Point (FP) calculation could be mapped into use-case and class diagram which have built before through the FP equation, and the results of the two diagrams can be analyzed following :

- The number of input application = 7
  Namely : login, logout, filling KRS, confirming KRS, updating KRS, checking the courses schedule, checking the courses value.
- The number of output application = 5
  Namely : KRS, subject, lecture schedule, lecturer, classroom.
- The number of online query = 7
  Namely : showing courses schedule, saving KRS, updating KRS, showing courses value, showing lecturers name, showing classroom, verify the username and password (login).
- The number of logic file = 5
  Namely : lecturer, courses value, subject, classroom, student.
- The number of external interface = 0

**Table 3. CFP Calculation of Academic System based on Object Oriented Model**

| Software System Components | Level of Complexity | | | | | | | | | Sum of CFP |
|---|---|---|---|---|---|---|---|---|---|---|
| | Low | | | Average | | | High | | | |
| | Count | Weighting Factor | Point | Count | Weighting Factor | Point | Count | Weighting Factor | Point | |
| | A | B | C = AxB | D | E | F = DxE | G | H | I = GxH | J = C+F+I |
| Input | 2 | 3 | 6 | 2 | 4 | 8 | 3 | 6 | 18 | 32 |
| Output | - | 4 | - | 4 | 5 | 20 | 1 | 7 | 7 | 27 |
| Online Query | 4 | 3 | 12 | - | 4 | - | 3 | 6 | 18 | 30 |
| Logic File | 2 | 7 | 14 | - | 10 | - | 3 | 15 | 45 | 59 |
| External Interface | - | 5 | - | - | 7 | - | - | 10 | - | - |
| Sum of CFP | | | | | | | | | | 148 |

Then calculate its RCAF :

**Table 4. RCAF Calculation of Academic System based on Object Oriented Model**

| No. | Subject | Value |
|---|---|---|
| 1 | The level of recovery reliability complexity | 0 1 2 3 4 (5) |
| 2 | The level of data communication complexity | 0 1 2 3 (4) 5 |
| 3 | The level of distributed processing complexity | 0 1 2 3 (4) 5 |
| 4 | Level of the need for performance complexity | 0 1 2 (3) 4 5 |
| 5 | The level of operating environment demand | 0 1 2 (3) 4 (5) |
| 6 | The level of developer knowledge needs | 0 1 2 3 (4) 5 |
| 7 | The level of updating the master file complexity | 0 1 2 3 4 (5) |
| 8 | The level of installation complexity | 0 1 (2) 3 4 5 |
| 9 | The level of input, output, online query and file application complexity | 0 1 2 3 (4) 5 |
| 10 | The level of data processing complexity | 0 1 2 3 4 (5) |
| 11 | The improbability level of reuse code | 0 1 2 3 (4) 5 |
| 12 | The level of customer organization variation | 0 (1) 2 3 4 5 |
| 13 | The extent of possible changes | 0 1 2 (3) 4 5 |
| 14 | Level of the ease of use demand | 0 1 2 3 (4) 5 |
| Total = RCAF | | 53 |

Then, the FP of academic system based on UML modeling can be calculated :
FP = CFP x ( 0.65 + 0.01 x RCAF )
= 148 x ( 0.65 + 0.01 x 53 )
= 174.64

## 5.2 Structural Modeling
- Entity Relationship Diagram

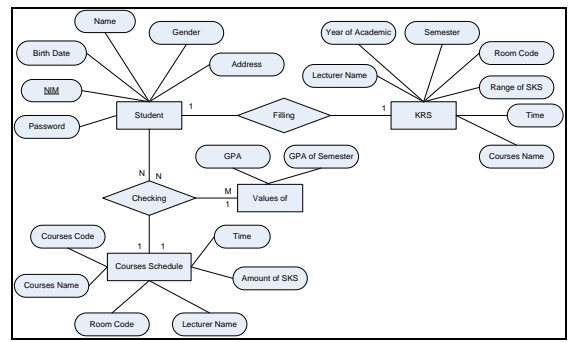

**Fig 4: ERD of Academic System**

- Data Flow Diagram

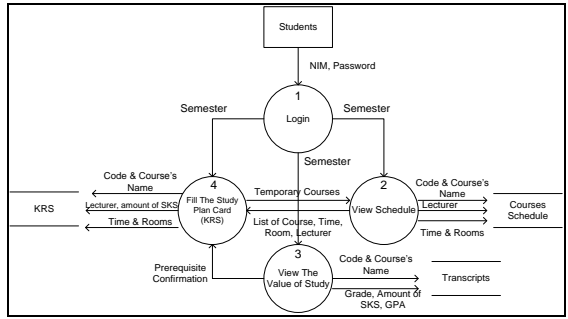

**Fig 5 : DFD of Academic System**

- Function Point Calculation

Same as in UML, function point calculation of structured model can be done by mapping ERD and DFD that have previously built into the formula of function point, and the results of the two diagrams can be analyzed as follows :

- The number of input application = 4
  Namely : login, filling KRS, checking the courses schedule, checking the courses value.
- The number of output application = 3
  Namely : KRS, courses schedule, transcripts
- The number of online query = 9
  Namely : login, showing courses schedule, filling KRS, checking KRS, showing courses value, showing lecturers name, showing classroom, checking transcript, updating KRS.
- The number of logic file = 6
  Namely : lecturer, courses value, subject, classroom, student, time
- The number of external interface = 0

**Table 5. CFP Calculation of Academic System based on Structural Model**

| Software System Components | Level of Complexity | | | | | | | | | Sum of CFP |
|---|---|---|---|---|---|---|---|---|---|---|
| | Low | | | Average | | | High | | | |
| | Count | Weighting Factor | Point | Count | Weighting Factor | Point | Count | Weighting Factor | Point | |
| | A | B | C = AxB | D | E | F = DxE | G | H | I = GxH | J = C+F+I |
| Input | - | 3 | - | 1 | 4 | 4 | 3 | 6 | 18 | 22 |
| Output | - | 4 | - | - | 5 | - | 3 | 7 | 21 | 21 |
| Online Query | 3 | 3 | 9 | - | 4 | - | 6 | 6 | 36 | 45 |
| File Logic | - | 7 | - | 3 | 10 | 30 | 3 | 15 | 45 | 75 |
| External Interface | - | 5 | - | - | 7 | - | - | 10 | - | - |
| Sum of CFP | | | | | | | | | | 163 |

**Table 6. RCAF Calculation of Academic System based on Structural Model**

| No. | Subject | Value | | | | | |
|---|---|---|---|---|---|---|---|
| 1 | The level of recovery reliability complexity | 0 | 1 | 2 | ③ | 4 | 5 |
| 2 | The level of data communication complexity | 0 | 1 | 2 | 3 | ④ | 5 |
| 3 | The level of distributed processing complexity | 0 | 1 | 2 | 3 | 4 | ⑤ |
| 4 | Level of the need for performance complexity | 0 | 1 | 2 | 3 | ④ | 5 |
| 5 | The level of operating environment demand | 0 | 1 | 2 | ③ | 4 | 5 |
| 6 | The level of developer knowledge needs | 0 | 1 | 2 | ③ | 4 | 5 |
| 7 | The level of updating the master file complexity | 0 | 1 | 2 | 3 | 4 | ⑤ |
| 8 | The level of installation complexity | 0 | 1 | ② | 3 | 4 | 5 |
| 9 | The level of input, output, online query and file application complexity | 0 | 1 | 2 | 3 | ④ | 5 |
| 10 | The level of data processing complexity | 0 | 1 | 2 | ③ | 4 | 5 |
| 11 | The improbability level of reuse code | 0 | 1 | 2 | 3 | ④ | 5 |
| 12 | The level of customer organization variation | 0 | 1 | ② | 3 | 4 | 5 |
| 13 | The extent of possible changes | 0 | ① | 2 | 3 | 4 | 5 |
| 14 | Level of the ease of use demand | 0 | 1 | 2 | ③ | 4 | 5 |
| | Total = RCAF | | | 46 | | | |

Then the function point of academic system based on structured modeling can be calculated :

$$FP = CFP \times (0.65 + 0.01 \times RCAF)$$
$$= 163 \times (0.65 + 0.01 \times 46)$$
$$= 180.93$$

## 6. CONCLUSIONS

From this paper can be taken several conclusions including:

1. Function point analysis method can serve as an alternative to calculating the volume of software system based on the complexity, both for the object-oriented model as well as a structured model.

2. The use of function point analysis method requires an experienced professional intervention because its very subjective calculations.

3. Due to the more calculation based on data processing representation, function point analysis method must also supported additional data to strengthen the estimated of software system volume that have been produced.

4. Function point that is produced for object-oriented and structured model not significantly different, where function point of object-oriented is 174.64 and function point of structured model is 180.93. So it can be stated that the use of object-oriented method or structure method quite good to give an idea has the estimation results of the process estimation.